# Emergence of pressure-induced metamagnetic-like state in Mn-doped CdGeAs$_2$ chalcopyrite

T. R. Arslanov[1a)], A. Yu. Mollaev[1], I. K. Kamilov[1], R. K. Arslanov[1], L. Kilanski[2], V. M. Trukhan[3], T. Chatterji[4], S. F. Marenkin[5] and I. V. Fedorchenko[5]

[1]*Amirkhanov Institute of Physics, Daghestan Scientific Center RAS 367003 Makhachkala, Russia*
[2]*Institute of Physics, Polish Academy of Sciences, Al. Lotnikow 32/46, 02-668 Warsaw, Poland*
[3]*Scientific-Practical Materials Research Centre (SSPA) of NAS of Belarus, 220072 Minsk, Belarus*
[4]*Institute Laue-Langevin, Boıˆte Postale 156, 38042 Grenoble Cedex 9, France*
[5]*Kurnakov Institute of General and Inorganic Chemistry RAS, 119991 Moscow, Russia*

The effect of hydrostatic pressure on resistivity and magnetic ac susceptibility has been studied in Mn-doped CdGeAs$_2$ room-temperature (RT) ferromagnetic chalcopyrite with two types of MnAs micro-clusters. The slight increase of temperature by about 30 K in the region between RT and Curie temperature $T_C$ causes a significant change in the positions of pressure-induced semiconductor-metal transition and magnetic phase transitions in low pressure area. By conducting measurements of the anomalous Hall resistance in the field $H≤5$ kOe, we present experimental evidence for pressure-induced metamagnetic-like state during the paramagnetic phase at pressure $P≈5$ GPa.

One of the most intensively studied topics both in condensed matter physics and from the viewpoint of functional materials are diluted magnetic semiconductors (DMS). Since more than 15 years the study of the relationship between transport and magnetic properties in DMS has been a widely debated issue [1-4]. In general, it should be noted that the possibility of the formation of long-range ferromagnetic order above room temperature in the prototype DMS (GaMn)As [5] does not find a complete explanation [6, 7]. Optimization of the growth parameters and postgrowth annealing procedure allows to reach a high Curie temperature ($T_C$) up to 190 K [8], however, is far from reaching 300 K.

Since 2000, the next generation of DMS compounds belonging to II-IV-V$_2$ compounds with chalcopyrite structure have been proposed (isovalent analogues of III-V semiconductors), due to the possibility of tailoring a new class of room temperature (RT) ferromagnetics [9]. Moreover, the primary experimental reports show the existence of RT ferromagnetism in ternary CdGeP$_2$:Mn [10], Zn$_{1-x}$Mn$_x$GeP$_2$ [11] and CdSnP$_2$:Mn [12] compounds. Thus, it is should induce their consideration as relevant materials for semiconductor spintronics [11]. High-resolution x-ray (XRD) method and NMR analysis showed that the RT ferromagnetism in II-IV-V$_2$:Mn is caused due to the presence of inhomogeneities and additional phases [12,13]. Based on these studies, it was suggested that high-temperature or RT ferromagnetism ($T_C$=320–355 K) in the II-IV-V$_2$:Mn is due to the presence of additional phases or clusters, but not intrinsic ferromagnetism (i.e. the case of pure DMS). In view of self-organizing clusters or second ferromagnetic phases in such hybrid semiconductor/metal systems, the potential effect of giant magnetoresistance (GMR) can find successful application in hard drives and memory storage devices. Beside of GMR these materials possess unique properties under pressure, which may extend them practical application as a pressure sensor.

Recently the studies of II-IV-V$_2$:Mn compounds under high hydrostatic pressure showed a number of interesting phenomena associated with anomalies of their magnetic and electrical properties [14,15]. In addition, the magnetovolume effect in Cd$_{1-x}$Mn$_x$GeAs$_2$ reaches a value of spontaneous magnetostriction $\omega_S$=0.5–1.7% that significantly shifts the $T_C$ with d$T_C$/d$P$ ≈ -14.0 ÷ -6.8 K/GPa, which greatly exceeds the value of 2.7 K/GPa for Ge$_{1-x}$Mn$_x$Te [16] and 1 K/GPa for (GaMn)As [17]. Thus, the ferromagnetic chalcopyrites demonstrate extreme sensitivity to external disturbances that motivates further interest to them via high-pressure studies.

In this letter, we carried out high pressure measurements of resistivity and the magnetic ac susceptibility with a weak increase in temperature above RT and below $T_C$ for CdGeAs$_2$ with different Mn content. The observed correlation between the magnetic and electrical properties is in our opinion due to the important contribution of MnAs clusters with hexagonal and rhombohedral structures. In addition we shown that the unusual increase of hysteresis loops of anomalous Hall resistance (AHR) in a field $H≤5$ kOe with increasing pressure is an evidence of the pressure-induced metamagnetic state [14]. We believe that the increase in AHR, and therefore the magnetization indicates to competing mechanism of interaction between MnAs clusters under pressure.

The bulk polycrystalline CdGeAs$_2$ samples doped with different Mn content were grown using a direct fusion method. The details of the growth procedure are presented in Ref. 18. According to XRD analysis of samples [19] the main crystal phase was identified as the tetragonal chalcopyrite structure CdGeAs$_2$ with space group $I\bar{4}2d$. The lattice parameters of all the crystals were close to pure CdGeAs$_2$ with $a$ = 5.942 (2) Å, $c$ =11.224 (2) Å [18] and changed monotonically ($a$ decreased while $c$ increased by around 0.2%) with increasing Mn content in the compounds. In addition to the main phase of the crystal, hexagonal and rhombohedral MnAs clusters are also detected. Scanning electron microscope studies revealed the presence of MnAs clusters with a diameter of about 1 μm in all the investigated samples. The chemical composition measured in different regions of the crystals shows that only a small fraction of Mn ions (around 1 at. %) are incorporated into the host semiconductor ($CdGeAs_2$). Figure 1(b) shows a schematic illustration of such composite system − $Cd_{1-x}Mn_xGeAs_2$ with MnAs clusters. A typical XRD patterns for $Cd_{1-x}Mn_xGeAs_2$ with x=0.18 are given in Fig. 1(a). The XRD patterns for all our samples indicate the presence of diffraction patterns associated to host $CdGeAs_2$ semiconductor, and two MnAs phases i.e. hexagonal and rhombohedral clusters [see the inset to Fig. 1(a)].

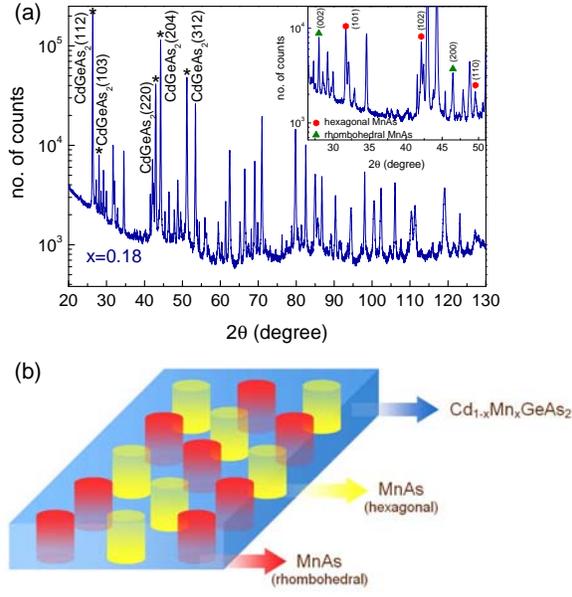

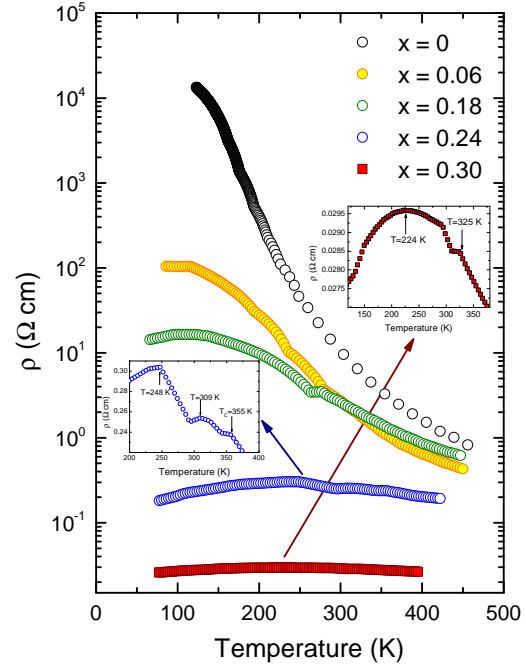

FIG 1. (Color online) The XRD patterns of $Cd_{1-x}Mn_xGeAs_2$ with x=0.18 (a). The inset shows magnified area of the diffraction pattern of sample in which the reflections from MnAs phases were visible. Schematic illustration of Mn doped chalcopyrite $CdGeAs_2$ with two types of MnAs ferromagnetic clusters (b).

FIG. 2. (Color online) The temperature dependence of resistivity for $Cd_{1-x}Mn_xGeAs_2$ samples with different Mn-contents. The insets represent an enlarged image of the transition from metallic to semiconductor type conductivity for samples x=0.24 and 0.30.

High hydrostatic pressure was generated in a Toroid type high-pressure cell [20]. A mixture of ethanol-methanol 4:1 was used as the liquid transmission medium. The magnetic ac susceptibility measurements were carried out on torus-shaped samples with the use of a frequency method integrated in a high pressure teflon capsule [15]. The electrical resistivity ($\rho$) at atmospheric pressure as well as high pressure was investigated by a standard four-probe method on samples with typical dimensions 3×1×1 mm$^3$. The pressure inside the capsule was continuously controlled by use of a manganin sensor. For measurements of Hall resistance ($R_{Hall}$), helical coil was used to create a magnetic field with a maximum of H=5 kOe.

Figure 2 shows the temperature dependence of the resistivity for $CdGeAs_2$ and $Cd_{1-x}Mn_xGeAs_2$ with x = 0.06–0.30 measured at ambient pressure. The $CdGeAs_2$ sample with increasing temperature shows a typical semiconductor behaviour. A change of the conductivity type, in particular for compositions with x=0.24 and x=0.30 with the increase in the Mn doping level is observed. At low temperatures (at $T<248$ K for x=0.24 and $T<224$ K for x=0.30) the metallic conductivity caused by the contribution of MnAs micro clusters dominated. At high temperatures ($T>300$ K) anomalies of the electrical properties were observed. In our opinion this features related with magnetic transitions of various MnAs clusters (insets to Fig. 2). It is noteworthy, that the singularity at $T=355$ K for x=0.24 corresponds exactly to the value of $T_C$ that is obtained on the basis of the magnetometeric measurements [18]. It should be noted for composition with x=0.24 in the near vicinity of $T_C$ the normal coefficient $R_0$ and the anomalous Hall coefficient $R_S$ related by $R/H=R_0+R_S(M/H)$ at $T>300$ K show a strong dependence on temperature [15].

From these measurements of $\rho(T)$ the presence of MnAs clusters contributes significantly to changes of conductivity type at higher levels of the Mn content in the alloys. However, in the case of the minimum Mn content (x=0.06÷0.18), these effects are not expressed, suggesting that the contribution to the conductivity of the substituted chalcopyrite $Cd_{1-x}Mn_xGeAs_2$ not essential. The description of $\rho(T)$ in such composite systems can be ambiguous, since it is necessary to consider different values of the mean free path ($\Lambda$), associated with different atomic configurations in the alloy [21]. On one hand it is a case of $Cd_{1-x}Mn_xGeAs_2$ where $\Lambda$ can be compared with the lattice constant $a$. On the other hand for various groups of MnAs micro clusters the value $\Lambda$<< long-range order of phase separation scale, which corresponds to a decrease $\rho$ at low Mn content in the composite. However, at high Mn content (x>0.24), the scale of phase separation tend to short-range order, and become <$\Lambda$, that implies an increase $\rho$.

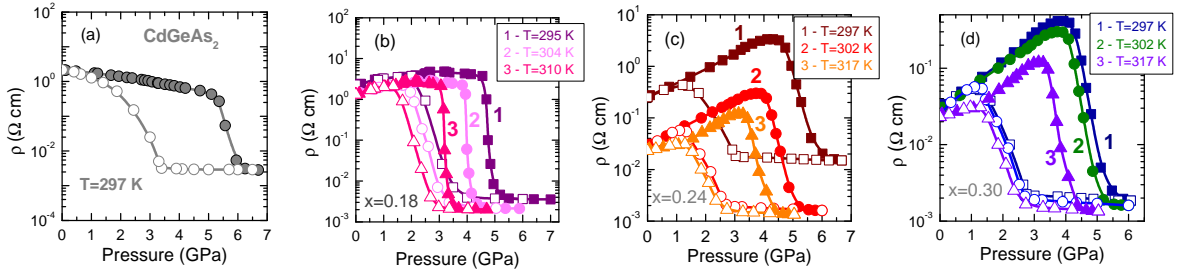

FIG. 3. (Color Online) High pressure dependences of resistivity at increase (solid symbols) and decrease of pressure (open symbols) for host semiconductor $CdGeAs_2$ at RT regime (a), $Cd_{1-x}Mn_xGeAs_2$ with x=0.18 (b), 0.24 (c) and 0.30 (d) samples at several temperatures.

Next, we studied pressure dependence of the transport properties of both the host semiconductor and $Cd_{1-x}Mn_xGeAs_2$ with x=0.18–0.30. Figure 3 presents a high pressure measurement of resistivity $\rho(P)$. The measurements were performed during increasing and decreasing the pressure. For all composition, during the decrease of the pressure, a hysteresis occurred, an evidence for restoration of the structural parameters after pressure cycle. As compared with prototype ferromagnetic chalcopyrite $Cd_{1-x}Mn_xGeP_2$, such hysteresis was observed only at a higher level of Mn doping [22]. Indeed, the effect of Mn doping introduces appreciable changes to the electrical transport. For compositions with x>0.18 [Fig 3 (b)–(d)], we observed an increase of the $\rho(P)$ with increasing pressure, while the host sample showed reduction. At high pressures for P>5 GPa for $CdGeAs_2$ noticeable pressure-induced semiconductor–metal transitions were found. The value of conductivity increases by almost 3 orders of magnitude, which probably attributed to structural changes [Fig 3(a)]. However, these transitions are the most pronounced and show maxima for x=0.24 and x=0.30. As for the sample with x=0.30 [Fig 3(d)] in the transition region (P=3.9 GPa and T=297 K) the value of the conductivity is $\sigma = 2.42$ $\Omega^{-1}cm^{-1}$ and after the transition at P=6 GPa reach $\sigma = 1000$ $\Omega^{-1}cm^{-1}$. The more important finding is associated with influence of temperature in the samples. A slight increase in temperature leads to a shift of phase transitions to the low-pressure and reducing the width of the hysteresis.

Figure 4 shows the pressure dependence of the magnetic ac susceptibility ($\chi_{AC}/\chi_0$ – susceptibility normalized to ambient pressure) at slight change of temperature regime up to $T_C$ of samples. In our previous studies we showed the pressure-induced responses of magnetic susceptibility at RT regime in $Cd_{1-x}Mn_xGeAs_2$ which were associated with both metamagnetism in the pressure range of 1.6–1.9 GPa and transitions to the paramagnetic phase at 4.1–4.7 GPa [14]. Here we note that the temperature effect noticeably affects on the sensitivity of peaks. Intensity of signal decreases and moves to the low pressure with P=1.6 GPa to P=0.65 GPa as at temperature increases by 32 K for sample with Mn content x=0.30 [FIG. 4(a)]. A similar behavior was observed for the sample with Mn content x=0.24 [FIG. 4(b)]. Under pressure the position of the metamagnetic transition $T_M$ sharply reacts to the temperature rise up to $T_C$ with $dT_M/dP \approx -33.7$ K/GPa for x=0.30 ($dT_M/dP \approx -25$ K/GPa for x=0.24). We believe that such a strong shift of the peaks rather corresponds to interaction between the MnAs clusters than intrinsic ferromagnetism of $Cd_{1-x}Mn_xGeAs_2$. However, the effect of temperature was weaker in the high-pressure area at P≥4 GPa (shifting the transition area about 0.5 GPa), where the transition to the paramagnetic phase occurred.

In general, we found a good correlation between the transport and magnetic properties under pressure. On the basis of the above observations, we can conclude that a perceptible shifts, as the peaks $\chi_{AC}(P)/\chi_0$ and hysteresis $\rho(P)$ (an average of 1 GPa when the T~10-15 K) in an area of low pressure is due to the relaxing response of MnAs clusters. Further, as a proof of the existence of pressure-induced metamagnetic-like state in Fig. 5 we show the magnetic field dependence of $R_{Hall}$, measured in a field $H \leq 5$ kOe at fixed pressures and RT. A very unusual behavior was found in two studied $Cd_{1-x}Mn_xGeAs_2$ samples with x=0.18 and x=0.30. With increasing pressure an enhancement in the hysteresis loops of AHR up to $P$=3.9 GPa occurs for x=0.18 [Fig 5(a)] and $P$=3.5 GPa for and x=0.30 [Fig 5(b)], respectively. However, a further increase in pressure instilled a typical paramagnetic behavior. The observed paramagnetic state at the high pressure area is in excellent agreement with $\chi_{AC}(P)/\chi_0$ data from Fig. 4. In addition, the transition from metamagnetic-like state to the paramagnetic state at $P$=4.8 GPa and $P$=5 GPa is accompanied by semiconductor-metal transition, as it follow from transport measurements (Fig. 3). It should be mentioned that the origin of the metamagnetic state can be explained as: (i) first-order transition between the antiferromagnetic and ferromagnetic states are induced by the magnetic field [23], (ii) from the realization of the two states with high and low spins in the MnAs [24].

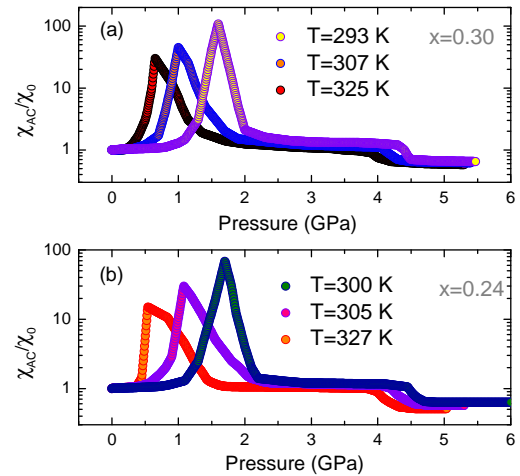

FIG. 4. (Color online) Pressure dependence of normalized magnetic ac susceptibility for $Cd_{1-x}Mn_xGeAs_2$ with x=0.30 (a) and 0.24 (b) at several temperature levels.

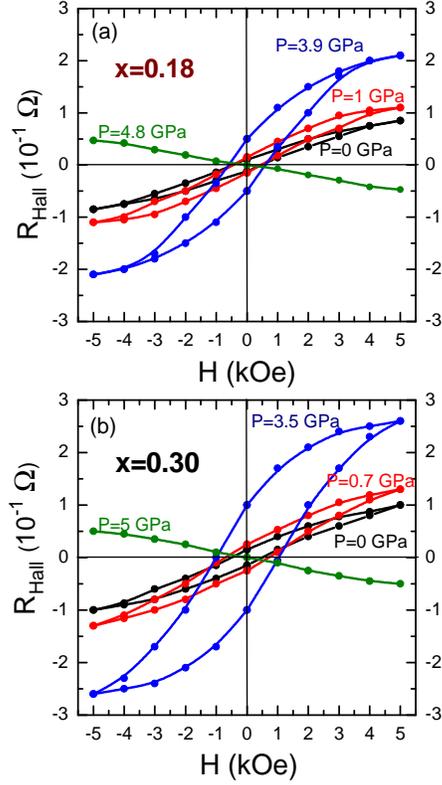

FIG. 5. (Color online) Hall resistance as a function of magnetic field for $Cd_{1-x}Mn_xGeAs_2$ with x=0.18 (a) and x=0.30 (b) at several fixed values of pressure and RT. Curves at $P$=4.8 GPa and $P$=5 GPa inverted relative to the Y axis.

In most cases of magnetic field induced metamagnetic transitions, the pressure effect leads to a significant decrease in transition temperature, $T_C$ [25, 26]. Similarly, the application of high pressure to bulk MnAs compound should lead to the disappearance of ferromagnetic state [24]. This in turn imposes controversy for case of MnAs clusters in Mn-doped $CdGeAs_2$ chalcopyrite. At the same time, appearance of a second ferromagnetic phase for bulk MnAs with the magnetic moment $3.24\mu_B$ at 0.6 GPa and $T$= 128 K is reported [27]. Moreover, the magnetic properties of the MnAs clusters are expected to be different than that of bulk MnAs [28,29]. In both cases, the presence of AHR feasibly, but the saturation tend of hysteresis loops at 5 kOe is typically for hybrid system (clusters containing) [30]. Thus, as we observed the pressure-induced metamagnetic-like behavior can be interpreted as the mechanism of competing interactions between the magnetic MnAs clusters with hexagonal and rhombohedral structures. The observed pressure-induced magnetic phase transitions indicate response from both groups of clusters with respect to pressure. Further study of the structural properties of Mn-doped chalcopyrite under high pressure synchrotron XRD will help obtaining a better understanding of the processes.

In summary, we have shown a good correlation between transport and magnetic properties under pressure (up to 6÷7 GPa) with weak change in temperature (~30 K). Contribution to the observed semiconductor-metal transition and magnetic phase transitions is more significant from MnAs micro clusters unlike contribution from substituted chalcopyrite $Cd_{1-x}Mn_xGeAs_2$. Confirmation for the existence of a pressure-induced metamagnetic-like behavior was shown in favor of which indicates an enhancement in the hysteresis loops of AHR in a wide range of pressures (up to ~4.8 GPa for x=0.18 and 5 GPa for x=0.30). Above these pressure areas occurs subsequent pressure-induced paramagnetic state.

This work was supported by the RAS Presidium Program №. 2 "Matters at High Energy Density", Section 2 "Matters Under High Static Compression", partly financed from funds for science in 2011-2014, under the Project No. N202 166840 granted by the National Center for Science of Poland, and partly supported by RFBR, research projects N 12-03-31203. A.T.R. thanks U. Z. Zalibekov for assistance at high pressure measurements and Artur Kowis for technical support.

FIG 1. (Color online) The XRD patterns of $Cd_{1-x}Mn_xGeAs_2$ with x=0.18 (a). The inset shows magnified area of the diffraction pattern of sample in which the reflections from MnAs phases were visible. Schematic illustration of Mn doped chalcopyrite $CdGeAs_2$ with two types of MnAs ferromagnetic clusters (b).

FIG. 2. (Color online) The temperature dependence of resistivity for $Cd_{1-x}Mn_xGeAs_2$ samples with different Mn-contents. The insets represent an enlarged image of the transition from metallic to semiconductor type conductivity for samples x=0.24 and 0.30.

FIG. 3. (Color Online) High pressure dependences of resistivity at increase (solid symbols) and decrease of pressure (open symbols) for host semiconductor $CdGeAs_2$ at RT regime (a), $Cd_{1-x}Mn_xGeAs_2$ with x=0.18 (b), 0.24 (c) and 0.30 (d) samples at several temperatures.

FIG. 4. (Color online) Pressure dependence of normalized magnetic ac susceptibility for $Cd_{1-x}Mn_xGeAs_2$ with x=0.30 (a) and 0.24 (b) at several temperature levels.

FIG. 5. (Color online) Hall resistance as a function of magnetic field for $Cd_{1-x}Mn_xGeAs_2$ with x=0.18 (a) and x=0.30 (b) at several fixed values of pressure and RT. Curves at *P*=4.8 GPa and *P*=5 GPa inverted relative to the Y axis.

a) Electronic mail: arslanovt@gmail.com